# A Microsoft Word Font for Anti-Matter

J. J. Reidy, L. N. Bolen, L. M. Cremaldi, T. J. Ray, D. A. Sanders, and D. J. Summers

Department of Physics and Astronomy

University of Mississippi-Oxford, University, MS 38677

**Abstract**

One often wishes to quickly add a few overlined characters such as $\bar{B}^0$ or $\bar{\nu}$ to a Microsoft Word document. <u>Underlined</u> characters are straightforward but overlined characters require equation editor which makes small picture files. The font here allows one to directly add overlined English and the most used overlined Greek characters to Microsoft Word documents on Apple Macintosh computers.

## Introduction

Several of the authors' particle physics experiments have involved anti-matter. Electromagnetic calorimeters [1,2] tested at SLAC have been used to measure $\bar{D}^0 \rightarrow K^+ \pi^- \pi^0$ decays at Fermilab [3] and $B^0$ - $\bar{B}^0$ oscillations [4] at BABAR. Work at CERN has examined $p\bar{p} \rightarrow B^0 - \bar{B}^0 X$, $B^0 \rightarrow \bar{B}^0 \rightarrow \mu^- \bar{\nu}_\mu X$ decays [5] and $\bar{p}$ annihilation in molybdenum [6]. Work at Fermilab has used kaon ID [7] to study $D^\pm$, $D^0$, and $\bar{D}^0$ production [8] and help reconstruct [9] $\bar{D}^0 \rightarrow K^+ K^-$ decays [10]. The overlined characters in this paragraph and in the references were generated by the font discussed in the next section.

## Overlined English and Greek Characters

The overlined English characters are produced in lower and upper case by the LinguistA 12 and LinguistA 24 fonts. A shift 5 is typed before typing the character. The shift 5 generates the overline rather than a percent sign. Without the shift 5, LinguistA generates plain English characters. The English fonts often look better in boldface. Note that the overline is actually a separate character and can sometimes become separated from the character below. LinguistA also includes the most common Greek letters that are used for anti-particles. Shift 5, option shift "v" gives an anti-neutrino, $\bar{\nu}$. Shift 5, option shift "z" gives an anti-Lambda, $\bar{\Lambda}$. For more Greek characters use "Key Caps" from the Apple Menu with the font set to LinguistA. Press the option and shift keys to see the Greek characters.

Here is a sample of the font.

LinguistA 12    ā b c̄ d̄ e f̄ ḡ h̄ ī j k̄ l̄ m n̄ ō p q̄ r s̄ t̄ ū v w̄ x ȳ z̄

LinguistA 12    a b c d e f g h i j k l m n o p q r s t u v w x y z

LinguistA 12    Ā B C̄ D̄ E F̄ Ḡ H̄ Ī J K̄ L̄ M N̄ Ō P Q̄ R S̄ T̄ Ū V W̄ X Ȳ Z̄

LinguistA 12    A B C D E F G H I J K L M N O P Q R S T U V W X Y Z

The fonts may be downloaded from

http://www.phy.olemiss.edu/HEP/LinguistA.sea.hqx

Drag and drop the font files onto your system folder. The file type should be recognized and the font will be put into your font folder. This font works in Microsoft Word under Apple MacOS [11, 12]. This study was supported by U. S. Department of Energy grant DE-FG05-91ER40622.